# Unexpected Spontaneously Dynamic Oxygen Migration

# on Carbon Nanotubes

Guangdong Zhu[1], Zhijing Huang[1], Liang Zhao[1*] and Yusong Tu[1*]

[1]College of Physical Science and Technology, Yangzhou University, Jiangsu, 225009, China

*Corresponding author: zhaoliang@yzu.edu.cn; ystu@yzu.edu.cn

## Abstract

Using the density functional theory calculations, we show that the oxygen functional groups exhibit unexpected spontaneously dynamic behaviors on the interior surface of single-walled carbon nanotubes (SWCNT). Two types of dynamic oxygen migrations - hydroxyl and epoxy migrations - are achieved by the breaking/reforming of C-O bond reaction and the proton transfer reaction. It is demonstrated that the spontaneously dynamic characteristic is attributed to the sharply reduced energy barrier less than or comparable to thermal fluctuations. We also observe a stable intermediate state with a dangling C-O bond, which permits the successive migration of oxygen functional groups. However, on the exterior surface of SWCNT, the oxygen groups are difficult to migrate spontaneously due to the relatively high energy barriers, and the dangling C-O bond prefers to transform into the more stable epoxy configuration. The spontaneous oxygen migration is further confirmed by the long-distance oxygen migration, which comprises three hydroxyl migration reactions and one C-O bond reaction. Our work provides a new understanding of the behavior of oxygen functional groups on interfaces and gives a potential route to design new carbon-based dynamic materials.

## Introduction

Understanding the behavior of functional groups at interfaces is of fundamental importance in physical, chemical, and biological processes such as the ions adsorption,[1-3] catalysis,[4-7] ligand recognition/binding,[8, 9] and protein folding,[10-12] as well as the focus of various potential applications ranging from functional material fabrication,[13-18] water purification[3, 19]/desalination[20-23] to drug synthesis and delivery,[24, 25] biosensor.[16, 26] In particular, oxygen functional groups such as epoxy, hydroxyl and carboxylic acid, covalently bond with the carbon skeleton of organic molecules and widely exist on the surface of carbon-based materials.[27-33] The presence of oxygen functional groups can enhance the interaction between the surface and surrounding solvents or absorbents, and thus alter the surface affinity,[34-36] hydrophobicity/hydrophilicity,[34, 37] diffusion resistance[38] and biocompatibility.[34] Despite various experimental and theoretical investigations over the past several decades, the behavior of oxygen functional groups on carbon-based material remains unclear.

Carbon nanotubes (CNTs), a carbon material by rolling-up single or multi-layers of graphene sheets, exhibit significant properties such as low density, high mechanical strength, excellent electrical and thermal conductivity, and have been a promising material as the building blocks of wide technical applications in the energy storage,[39, 40] mass transportation,[41-44] signal processing,[45-48] biomolecules manipulation,[49] alcohol/water separation[50] and nanoelectronic devices manufacturing.[51, 52] The covalently oxygen functional groups on the surface of CNT improve their dispersity in polar solvents.[53, 54] More importantly, they provide coordinate sites for extensive interactions between the interface and nanoparticles/clusters and macromolecules,[55-57] even catalyze the chemical reaction.[58] Generally, the oxygen functional groups on the CNT surface seldom migrate or change under ambient conditions. If the oxygen functional groups can migrate along the CNT surface, it is expected that CNT will show structural adaptivity during the interaction with molecules, clusters or particles, and the relevant applications such as biosensors and catalysts could also be benefited.

In this work, we have performed the density functional theory (DFT) calculations to study the dynamic behavior of oxygen functional groups on the interior and exterior surfaces of the single-walled carbon nanotube (SWCNT). Two typical oxygen migrations, the hydroxyl migration and the epoxy migration, are proved in detail. We also design a long-distance oxygen migration pathway on the interior surface of SWCNT. To the best of our knowledge, this is the first report of spontaneously dynamic oxygen migration on the interface of CNT.

**Methods**

The DFT calculations via the Gaussian 09 software package[59] are performed to analyze the reaction pathways of the oxygen migration. All geometry configurations are optimized to reach the convergence using the B3LYP[60, 61] functional with the 6-31G(d) basis set. The frequency analysis of normal vibrational model is carried out to check the validity of the transition state (TS). The solely one imaginary frequency affirms the true TS structure. Further, the TS is connected to both the reactants and products through the intrinsic reaction coordinate (IRC) calculation using the same level of theory.

**Results**

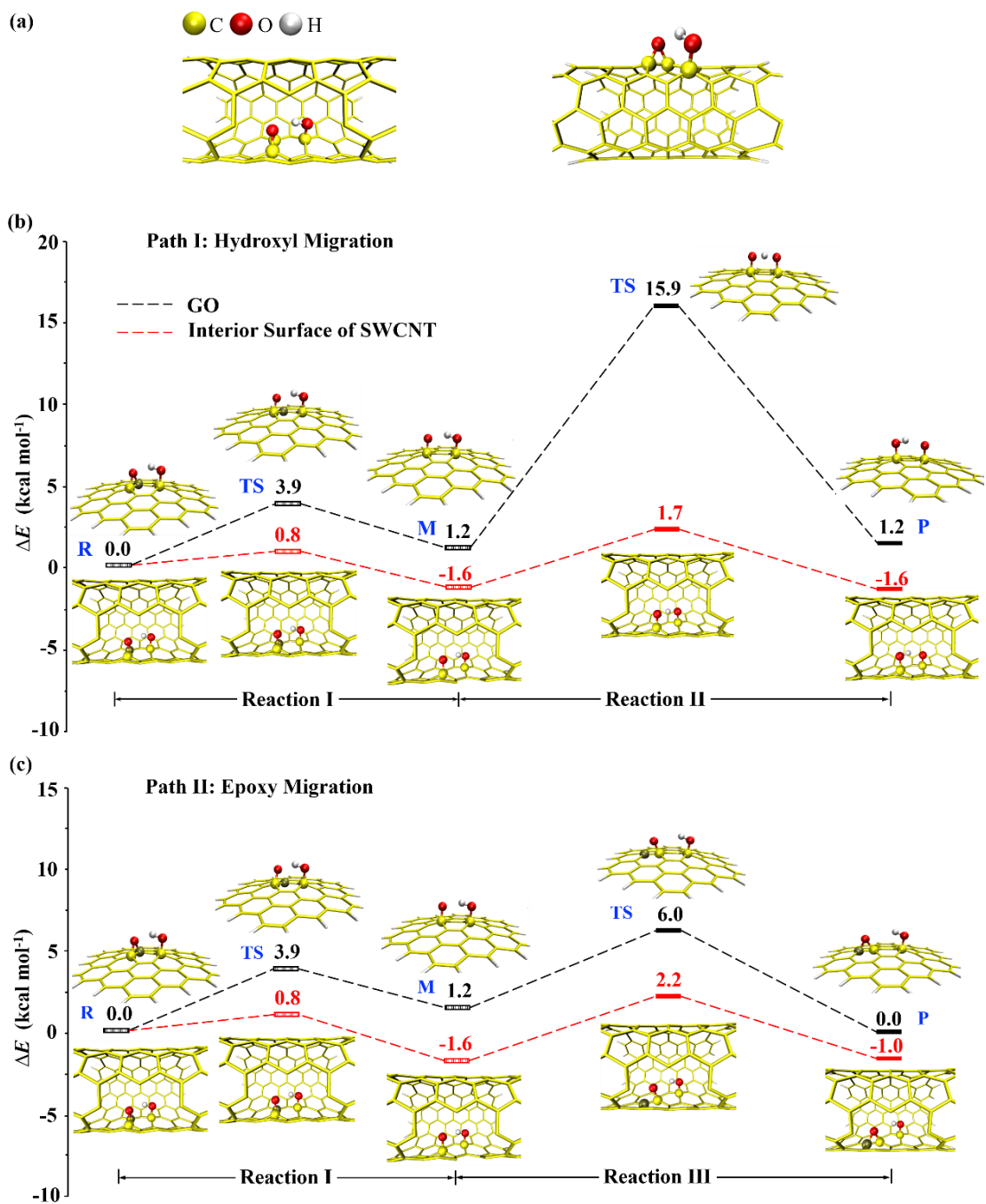

**Figure 1**. (a) Representative configurations of armchair (6, 6)-SWCNT with epoxy and hydroxyl groups on the interior and exterior surfaces. Pathways of two types of oxygen migrations - (b) hydroxyl migration and (c) epoxy migration - on GO (black lines) and the interior surface of SWCNT (red lines). Notation: reactants (R), intermediates (M), transition states (TS) and products (P); C-O bond breaking (Reaction I), proton transfer between the neighboring epoxy and hydroxyl for the exchange (Reaction II) and C-O bond reforming (Reaction III). The length of (6, 6)-SWCNT is 12.3 Å. The hydrogen atoms are added to saturate the edge carbon atoms. Carbon atoms in front of oxygen atoms on the SWCNT are hidden for the clarity of the migrations of hydroxyl and epoxy groups. The energy values of R for GO and SWCNT are all shifted to 0 for easy

comparison.

Figure 1(a) shows the representative configurations of armchair (6, 6)-SWCNT with dominated oxygen functional groups on the interior and exterior surfaces. A pair of epoxy and hydroxyl groups are introduced to mimic the correlated distributions of oxygen functional groups on the carbon-based materials.

Figure 1(b) illustrates the pathways of hydroxyl migration assisted by epoxy on GO and the interior surface of SWCNT. According to previous DFT calculations,[62] the hydroxyl migration on GO is assisted by epoxy through two successive reactions: the C-O bond breaking (Reaction I) and the proton transfer between the neighbouring epoxy and hydroxyl for the exchange (Reaction II). For Reaction I, the energy barrier on GO is 3.9 kcal/mol,[62] while the barrier is much lower, even down to 0.8 kcal/mol on the interior SWCNT surface. This indicates that the C-O bond breaks more easily on the interior surface of SWCNT than on GO. More surprisingly, for Reaction II, the energy barrier of proton transfer is significantly decreased to 3.3 kcal/mol, compared to 14.7 kcal/mol on GO. Since the energy barriers for Reactions I and II are all reduced to the order of thermal fluctuations on the interior surface of SWCNT, the hydroxyl can spontaneously migrate through the successive C-O bond breaking and proton transfer. The stability of intermediate (M) on the interior surface of SWCNT is also counterintuitive. Generally, the dangling C-O bond is less stable than the epoxy configuration as the latter with two C-O bonds. This is the case for M on GO, where the energy of M is 1.2 kcal/mol higher than R. However, on the interior surface of SWCNT, the M is found to be 1.6 kcal/mol lower than R, indicating that the dangling C-O bond can be stabilized.

Figure 1(c) shows the epoxy migration assisted by hydroxyl on GO and the interior surface of SWCNT. This process consists of the C-O bond breaking (Reaction I) and reforming reactions (Reaction III). The C-O bond breaking is same as that in the hydroxyl migration. The energy barrier of C-O bond reforming on the interior surface of SWCNT is 3.8 kcal/mol, still lower than 4.8 kcal/mol on GO. These relatively low energy barriers indicate that the epoxy can migrate spontaneously and easily on the interior surface of SWCNT. Besides, the intermediate is again stable as its energy is 1.6 kcal/mol and 0.6 kcal/mol lower than the R and P state, respectively.

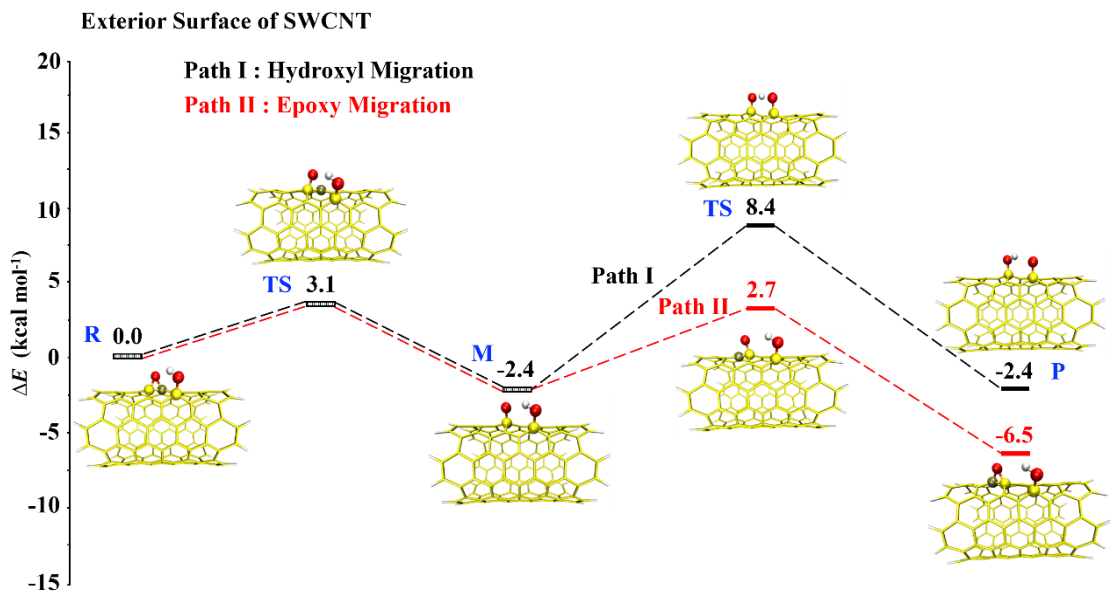

**Figure 2**. Oxygen migration including the hydroxyl migration (Path I, black lines) and the epoxy migration (Path II, red lines) on the exterior surface of SWCNT. The energy of R is shifted to 0 for easy comparison.

We further analyzed the two types of dynamic oxygen migrations - hydroxyl migration and epoxy migration - on the exterior surface of SWCNT. As shown in Fig. 2, the C-O breaking has a barrier of 3.1 kcal/mol. However, there remains a relatively high barrier of 10.8 kcal/mol for proton transfer in the hydroxyl migration and a moderate barrier of 5.1 kcal/mol for the C-O bond reforming reaction during the epoxy migration. These barriers for two types of oxygen migrations suggest that the epoxy can migrate more easily than hydroxyl. For the epoxy migration, the energy of the P state is 4.1 kcal/mol lower than that of the M state, implying that the dangling oxygen atom tends to transform to the epoxy configuration under thermal fluctuations.

The spontaneously dynamic oxygen migration and the existence of a stable intermediate on the interior surface of SWCNT might be universal. We analyzed the oxygen migration pathways for another size of armchair SWCNT, (5, 5)-type, while keeping all the other settings unchanged (See PS. 1 in the Supporting Information). On the interior surface of SWCNT, the energy barriers of C-O bond breaking reaction, proton transfer reaction between the neighboring hydroxyl for the exchange, and C-O bond reforming reaction are 1.1 kcal/mol, 3.8 kcal/mol and 6.5 kcal/mol, respectively. These values are totally the order of thermal fluctuations, indicating that the epoxy and hydroxyl groups can spontaneously migrate. Again the intermediate with a dangling C-O bond is found to be stable than the configuration of epoxy. On the exterior surface of SWCNT, it is difficult for the hydroxyl and epoxy to migrate spontaneously since there remains a high barrier of 11.9 kcal/mol for the proton transfer reaction and a moderate barrier of 6.9 kcal/mol for the C-O reforming reaction.

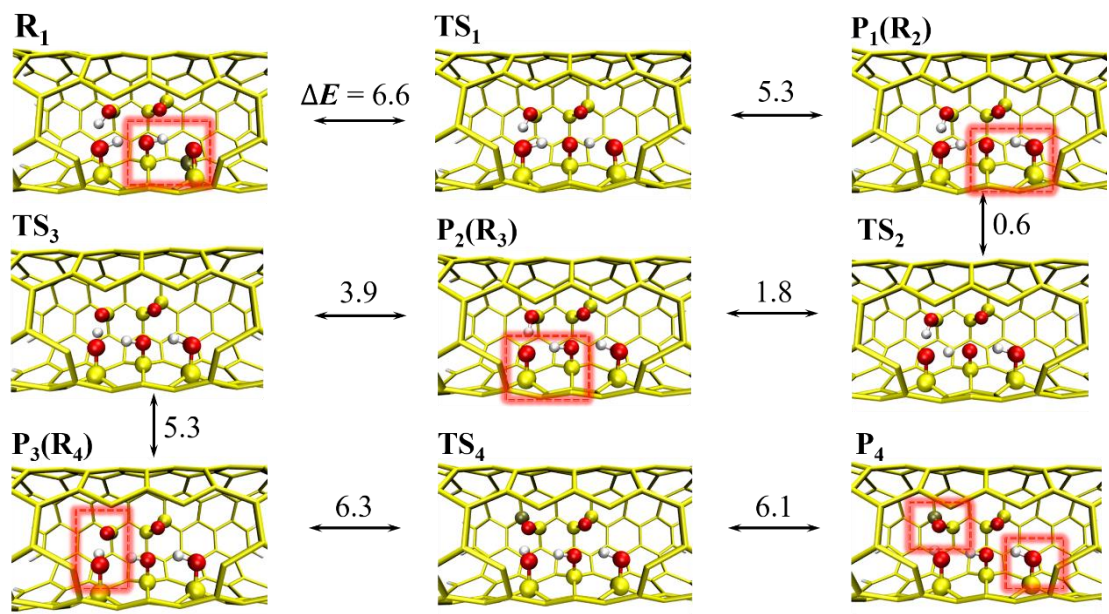

**Figure 3**. Oxygen migration over a long distance on the interior surface of (5, 5)-SWCNT through successive three hydroxyl migration reactions ($R_1$ to $P_1$, $R_2$ to $P_2$, and $R_3$ to $P_3$) and a C-O bond reforming reaction ($R_4$ to $P_4$). The red dashed box shows the oxygen functional groups involved in the reactions. The energy barriers $\Delta E$ are in the unit of kcal/mol.

The spontaneous oxygen migration can be further confirmed by the long-distance migration process associated with a successive process of three hydroxyl migration reactions and a C-O bond reforming reaction. Figure 3 demonstrates the reaction pathway of oxygen migration of epoxy and hydroxyl groups on the interior surface of (5, 5)-SWCNT. In the configuration from $R_1$ to $P_1$, a hydroxyl migrates toward the right direction, accompanied by the C-O bond breaking reaction and proton transfer reactions. With the emergence of an intermediate state in $P_1$, the hydroxyl continues migrating via two successive proton transfer reactions from $R_2$ to $P_2$ then from $R_3$ to $P_3$. At the final step from $R_4$ to $P_4$, the C-O bond reforms and we can see that the oxygen achieves a long-distance migration. The energy barriers in all reactions are all less and comparable to thermal fluctuations, indicating that the oxygen functional groups can spontaneously migrate over a long distance on the interior surface of SWCNT.

## Conclusions

In summary, we have found the unexpected spontaneously dynamic oxygen migration on the interior surface of SWCNT. The hydroxyl migration and epoxy migration are two types of dynamic oxygen migrations, which are achieved by the breaking/reforming of the C-O bonds reaction and the proton transfer reaction. This is attributed to the sharply reduced energy barrier less than or comparable to the thermal fluctuations. It should be noted that the migration of oxygen functional groups has already been theoretically and experimentally validated on the basal plane of GO.[62] However, only with the adsorption of water molecules can the oxygen functional groups spontaneously migrate, and the GO will be converted into a dynamic covalent material.

In experiments, the CNT has been demonstrated as an excellent nanoscale host for metal clusters and biomolecules[55, 63, 64]. It is expected that this innately and spontaneously dynamic characteristic of CNT can provide structural adaptivity for the response to the molecule adsorption to the interior surface. Especially for the metal clusters used in catalyzing the chemical reactions within the CNT,[58, 65] the structural adaptivity will improve the reaction rates and facilitate the reaction conditions. Therefore, our finding is fundamentally essential for designing dynamic materials, such as CNT-based nanoreactors with high reactivity and sensitivity.

Furthermore, the stable intermediate state with a dangling C-O bond is observed, which permits the C-O bond breaking/reforming and proton transfer reactions, and ensures the oxygen migration on the interior surface of SWCNT. It should be noted that the intermediate state has also been found on GO. However, the dangling C-O bond prefers to transform into the epoxy configuration. This finding thus provides new insights in understanding the chemical structures and behaviors of oxygen functional groups on the surface of CNT.

## Acknowledgment

We thank the helpful suggestions given by Drs. Zonglin Gu and Shuming Zeng. This work was supported by the National Natural Science Foundation of China [grant numbers 11605151, 12075201, 11675138]; Natural Science Foundation of Jiangsu Province. [Nos: BK20161325, BK20201428]; Special Program for Applied Research on Supercomputation of the NSFC‐Guangdong Joint Fund.